\documentclass{ws-procs975x65}

\begin{document}

\title{Quasilocal Center-of-Mass for Teleparallel Gravity}

\author{James M. Nester}

\address{Department of Physics and Institute of Astronomy,\\
National Central University,
Chungli, Taiwan 320, R.O.C.\\
E-mail: nester@phy.ncu.edu.tw}

\author{Fei-Hong Ho and Chiang-Mei Chen}

\address{Department of Physics,\\
National Central University,
Chungli, Taiwan 320, R.O.C.\\
E-mail:  cmchen@phy.ncu.edu.tw}

\maketitle

\abstracts{Asymptotically flat gravitating systems have 10
conserved quantities, which lack proper local densities. It has
been hoped that the teleparallel equivalent of Einstein's GR
(TEGR, aka GR${}_{||}$) could solve this gravitational
energy-momentum localization problem. Meanwhile a new idea:
quasilocal quantities, has come into favor. The  earlier
quasilocal investigations focused on energy-momentum. Recently we
considered quasilocal angular momentum for the teleparallel theory
and found that the popular expression (unlike our
``covariant-symplectic'' one) gives the correct result only in a
certain frame. We now report that the center-of-mass moment, which
has largely been neglected, gives an even stronger requirement. We
found (independent of the frame gauge) that  our ``covariant
symplectic'' Hamiltonian-boundary-term quasilocal expression
succeeds for all the quasilocal quantities, while the usual
expression cannot give the desired center-of-mass moment.  We also
conclude, contrary to hopes, that the teleparallel formulation
appears to have no advantage over GR with regard to localization.}

\section{Introduction and Overview}\label{sec:introverview}

Associated with the symmetries of flat Minkowski spacetime are 10
conserved quantities: energy-momentum, angular momentum and the
(often overlooked) center-of-mass moment. Asymptotically flat
gravitating systems also have these 10 conserved quantities. The
total quantities for the whole space are well defined. However,
unlike the situation for all the other matter and interaction
fields which constitute its source, the gravitational field itself
lacks {\it proper local densities} for these quantities. This can
be understood in terms of the equivalence principle: the
gravitational field cannot be detected at a point. The
localization of gravitational energy-momentum still remains one of
the outstanding problems in classical gravity theory.

In 1961 M{\o}ller proposed a tetrad-teleparallel\cite{Mol}
equivalent of Einstein's GR (TEGR, aka GR${}_{||}$) as a way to
solve this problem; this promising approach is still being
pursued.\cite{Ne89,Mal,AP,AGP,Mal2}

More recently a new idea, quasilocal quantities\cite{SzaLR} (see
Brown and York's seminal paper\cite{BY} for references to the
early work), has come into favor. Our approach has been to develop
a {\it covariant Hamiltonian formalism}.  In Hamiltonian
approaches quasilocal quantities are associated with the
Hamiltonian boundary term. This boundary term has, at least
formally, considerable freedom, which can be related to the choice
of boundary conditions. Certain principles can then be used to
select ``good'' boundary conditions.  In particular we have
advocated a simple, appropriate and reasonable choice to limit
this freedom: we found there are only {\bf two} {\it
covariant-symplectic} choices for each dynamic
field.\cite{CNT,CN9,CNC9,CNC0,CN0}

Here we consider the various proposed Hamiltonian boundary terms
for the quasilocal quantities of tetrad-teleparallel gravity.
Earlier investigations treated only
energy-momentum.\cite{Ne89,AP,AGP,Mal} Then Vu\cite{Vu} considered
angular momentum; he found that the popular expression (unlike our
``covariant-symplectic'' one) gives the correct result only in a
certain frame.  More recent TEGR angular momentum
results\cite{Mal2} are consistent with this conclusion.

The long neglected center-of-mass moment, however, offers the most
restrictive requirements, not only on the allowed asymptotic
behavior of the variables\cite{RT,BO,Sza} but also on the
acceptable form of the expressions. We tested the various
tetrad-teleparallel Hamiltonian boundary quasilocal expressions on
the asymptotic eccentric Schwarzschild solution. We found (frame
gauge independently) that our ``covariant symplectic''
Hamiltonian-boundary-term teleparallel quasilocal expression
succeeds for all of the quantities, while the usual {\it tetrad}
expression does not give the desired center-of-mass
moment.\cite{Ho}

It turns out that one of our ``good'' teleparallel expressions is
 equivalent to a standard ``covariant-symplectic'' quasilocal
Riemannian GR expression\cite{CN9,CN0} (the others are
asymptotically equivalent).  We conclude that, for localization
purposes, GR${}_{||}$ is {\it not} better than GR.  (This
assessment would be revised if a good gauge condition for
teleparallel GR is established.)

\section{Background}\label{sec:background}

 In {\em Newtonian physics}
Galilean symmetry leads to the  associated conserved quantities:
energy, momentum, angular momentum (AM) and the (often overlooked)
center-of-mass moment (COM). (The latter is associated with the
lack of a preferred inertial reference frame.) Likewise in {\em
special relativity}, Poincar{\'e} symmetry leads to conserved
quantities for particles: spacetime translations are associated
with energy-momentum, $P_\mu$, and Lorentz transformations
(including both boosts and rotations) are associated with
4-dimensional angular momentum: $L^{\mu\nu}:=x^\mu P^\nu- x^\nu
P^\mu$, which
 inseparably includes the covariant partners:
angular momentum, $L^{ij}=x^i P^j-x^j P^i$,   and  the {\em
center-of-mass} moment,
\begin{equation}
L^{0k}=x^0 P^k-x^k P^0\equiv ct P^k-x^k E/c.
\end{equation}
In classical field theory, Noether's theorem gives conserved
current local densities for energy-momentum (EM) and AM/COM; the
latter also includes, in general, a spin density term:
$J^{\mu\nu}=L^{\mu\nu}+S^{\mu\nu}$.\cite{Ne4}

\section{Gravitational Energy-Momentum and its
Localization}\label{sec:geml}

A basic notion of energy-momentum is the Noether conserved
quantity associated with the spacetime translation symmetry. This
idea works well enough in flat space---aside from the usual
conserved current ambiguity of  adding to the density a quantity
with identically vanishing divergence. The case of conserved
quantities for gravitating systems, however, is quite different.
It is well known that asymptotically flat spacetimes have well
defined total conserved quantities, whereas localization is
problematical.\cite{Ne4,MTW}

Note that the source of gravity is the energy-momentum density for
matter and for all other interaction fields.   Via this relation
gravity absolutely identifies the source energy-momentum density,
removing thereby the uncertainty in the conserved Noether
translational current.
%(Thus gravity can be used to detect the
%presence of otherwise undetected energy-momentum, e.g. dark
%matter).
It seems rather ironical that the energy-momentum density for
gravity itself is not so sharply defined.

Energy-momentum should be conserved. In view of the fact that it
is exchanged {\it locally} between the sources and gravity,
investigators sought some kind of local description of
energy-momentum for the gravitational field itself.  Standard
arguments (e.g., Noether symmetry), however, lead only to
energy-momentum densities which are reference frame dependent {\it
pseudotensors}.\cite{Ne4} It seems that the gravitational field
(hence any gravitating system, and consequently every {\em
physical} system) has no proper energy-momentum density (similarly
there is no proper angular momentum/center-of-mass density). This
can be understood in terms of the equivalence principle: gravity
cannot be detected at a point.\cite{MTW}

\section{The Tetrad-Teleparallel Approach}\label{sec:Mollertettelep}

In 1961 M{\o}ller proposed a certain reformulation of Einstein's
GR as a means of solving this localization problem.\cite{Mol}  On
the one hand it can be regarded simply as GR expressed in terms of
an orthonormal frame. Alternately it can be described in terms of
a different framework: {\it teleparallel geometry} (aka absolute
parallel, Weitzenb{\"o}ck geometry).  It then has a different
connection, a different parallel transport, which is a kind of
opposite to the Riemannian one: torsion is generally non-zero
while curvature vanishes (hence parallel transport is path
independent, but non-trivial). Common names for this theory are
TEGR (the teleparallel equivalent of GR) and GR${}_{||}$. A major
source of interest is that the energy-momentum density for this
theory is determined by an object which is {\it tensorial} (unlike
the usual GR description).  More precisely it is a tensor under
coordinate transformations---however it does depend on the Lorentz
gauge choice for the orthonormal frame. Hence to fix the
localization a frame gauge condition is necessary.  Although, as
far as we know,
 no satisfactory gauge condition has yet been recognized, this tetrad-teleparallel
approach is still regarded as promising and has been under active
investigation.\cite{Ne89,Mal,AP,AGP,Mal2}

\section{Quasilocal Energy-Momentum}\label{sec:qem}

With regard to the energy-momentum localization ``problem'', it is
now widely believed that the proper idea is not local but rather
{\it quasilocal} quantities\cite{SzaLR} (i.e., quantities
associated with a closed 2-surface).

\subsection{Covariant Hamiltonian Approach}

One approach to energy-momentum is via the Hamiltonian.  Energy
can be regarded as the value of the Hamiltonian.  The traditional
Hamiltonian techniques have certain virtues but they exact a
price: they are not manifestly covariant.  We have developed an
alternate {\it covariant Hamiltonian formalism}, which naturally
yields manifestly 4-covariant expressions.\cite{CN9,CN0} Here we
briefly review the key features.

We start with a first order Lagrangian for an f-form field,
\begin{equation}
{\mathcal L}=d\varphi\wedge p-\Lambda(\varphi,p),
\end{equation}
(independent variation with respect to $\varphi$ and $P$ yields
pairs of {\it first order} equations). From it we construct the
Hamiltonian 3-form (density)
\begin{equation} {\mathcal H}(N)=\pounds_N\varphi\wedge
p-i_N{\mathcal L},
\end{equation}
which is conserved ($d{\mathcal H}(N)=0$ ``on shell'').
Consequently its spatial integral yields a conserved quantity for
each choice of displacement $N$. A short calculation shows that
the Hamiltonian 3-form can be expressed in the form
\begin{equation}
{\mathcal H}(N)=N^\mu{\mathcal H}_\mu+d{\mathcal B}(N).
\end{equation}
For a (finite or infinite) spatial region, the Hamiltonian---the
integral of this 3-form over a spacelike hypersurface---is the
generator of field displacements along the vector field $N$.  The
total differential term (via the generalized Stokes theorem)
yields a  flux integral over the closed 2-surface boundary.

For all locally diffeomorphic invariant theories the spatial
surface density ${\mathcal H}_\mu$
 is proportional to field equations;
its value vanishes ``on shell''.   Hence the (conserved) value of
the Hamiltonian,
\begin{equation}
H(N):=\int_\Sigma {\mathcal H}(N)=\int_\Sigma N^\mu{\mathcal
H}_\mu +\oint_{\partial\Sigma} {\mathcal B}(N)\simeq
\oint_{\partial\Sigma} {\mathcal B}(N),
\end{equation}
is {\it quasilocal}, being determined by the boundary term
${\mathcal B}(N)$.

In order to have a proper Hamiltonian for the desired phase space
variables, the boundary term inherited from the Lagrangian can
(and indeed in general must) be adjusted (this was first nicely
explained for GR by Regge and Teitelboim;\cite{RT} since then the
arguments have been refined\cite{BO,Sza}). Note that the boundary
term, and hence the value of the Hamiltonian quasilocal
quantities, can be adjusted without changing the field equations
or the conservation property. The freedom in the Hamiltonian
boundary term is linked (via the symplectic structure of the
boundary term in the Hamiltonian variation) with the freedom of
choice of boundary conditions.

We found that for each dynamic field the boundary term naturally
inherited from the Lagrangian, ${\mathcal B}(N)=i_N\varphi\wedge
p$, has only two alternate replacements  which have
``covariant-symplectic'' Hamiltonian boundary variation terms.
They are
\begin{equation}
{\mathcal B}_\varphi(N) := i_N\varphi\wedge\Delta
  p-\varepsilon\Delta\varphi\wedge
  i_N{\buildrel \scriptstyle \circ \over {p}},
\end{equation}
\begin{equation}
{\mathcal B}_p(N) := i_N{\buildrel \scriptstyle \circ \over
{\varphi}}\wedge\Delta p-\varepsilon\Delta\varphi\wedge i_N p,
\end{equation}
where $\varepsilon=(-1)^f$, ${\buildrel \scriptstyle \circ \over
{\varphi}}$ and ${\buildrel \scriptstyle \circ \over {p}}$ are the
values in a reference configuration, and
$\Delta\varphi:=\varphi-{\buildrel \scriptstyle \circ \over
{\varphi}}$, $\Delta p:=p-{\buildrel \scriptstyle \circ \over
{p}}$. These boundary expressions, respectively, correspond to
Dirichlet and Neumann
 boundary conditions. Asymptotically, at spatial infinity,
both give the same values. For our considerations here we need
only consider one.

\section{Tetrad Gravity}\label{tetgrav}

By tetrad gravity we mean theories where the only dynamic variable
is the frame (sometimes referred to as a {\it tetrad} or {\it
vierbein}).  Technically for our purposes it is most convenient to
work with the coframe $\vartheta^\mu$ (i.e. the basis one-forms,
dual to the vector basis $e_\alpha$).  Here we shall confine our
attention to {\it orthonormal} frames. (The metric is then given
by $g=g_{\alpha\beta}\vartheta^\alpha\otimes\vartheta^\beta$ with
the orthonormal metric components
$g_{\mu\nu}=\hbox{diag}(-1,+1,+1,+1)$).

Any tetrad theory can be derived from a first order Lagrangian of
the form
\begin{equation}{\mathcal L}_{\rm tet}=d\vartheta^\mu\wedge\tau_\mu-
\Lambda_{\rm tet}(\vartheta,\tau).\label{tetradL}\end{equation}
The orthodox choice for $\Lambda_{\rm tet}$ is an expression
quadratic in $\tau$; then $\tau$ and $d\vartheta$ are linearly
related. For a certain choice, $\Lambda_{\rm tet}=\Lambda_{{\rm
GRtet}} $ (the specific expression is not needed here, a neat
version with an extra field is given in Eq. (\ref{TEGR}) below),
this gives GR in terms of a tetrad. It should be remarked that our
formalism here applies to {\it all} tetrad theories. The {\it
generic} tetrad theory\cite{HS} is a theory for a preferred
orthonormal frame with no local Lorentz gauge freedom. Only one
special (albeit highly interesting) case---the tetrad version of
GR---has local Lorentz gauge freedom.

The Hamiltonian boundary term obtained from the tetrad Lagrangian
(\ref{tetradL}) is
\begin{equation}
{\mathcal B}_{\rm
tet}(N)=i_N\vartheta^\mu\tau_\mu=N^\mu\tau_\mu.\label{Btet}
\end{equation}
We also consider one of our ``covariant-symplectic'' alternatives:
\begin{equation}
{\mathcal B}_\vartheta(N) := i_N\vartheta^\mu\Delta \tau_\mu
+\Delta\vartheta^\mu\wedge i_N{\buildrel \scriptstyle \circ \over
\tau}_\mu. \label{Btheta}
\end{equation}
This is just a slight refinement, reducing to (\ref{Btet}) if the
reference values are trivial (a natural choice).

Note in particular that the boundary terms given above have the
same form for all tetrad theories, independent of the specific
form of $\Lambda_{\rm tet}$. It is worth recalling here that GR,
in its Riemannian formulation, has {\it many} proposed
energy-momentum expressions; there is {\it no} consensus as to
which is the best. In contrast, tetrad
investigators\cite{Mol,Ne89,Mal,AP,AGP,Mal2} basically agree on
the expression (\ref{Btet}). The general tetrad theory has an
essentially {\it unique} energy-momentum flux expression, which
naturally applies to the special case GRtet.

\section{Metric-Compatible Gravity}\label{sec:mcg}

We now take a more geometric approach. We start with the general
class of geometries with a metric-compatible
connection.\cite{GH,B} One of our basic variables is the
orthonormal coframe $\vartheta^\alpha$. The other basic variable
is an {\it a priori independent} metric-compatible connection
one-form: $ \Gamma^{\alpha\beta}=\Gamma^{[\alpha\beta]}$. These
``potentials'' determine the
  {\it curvature 2-form}:
\begin{equation}
R^\alpha{}_\beta := d\Gamma^\alpha{}_\beta +
\Gamma^\alpha{}_\gamma\wedge\Gamma^\gamma{}_\beta,
\end{equation}
and the {\it torsion 2-form}:
\begin{equation}
T^\alpha := D\vartheta^\alpha := d\vartheta^\alpha +
\Gamma^\alpha{}_\beta\wedge\vartheta^\beta.
\end{equation}

The general metric-compatible  geometric first order Lagrangian
has the form
\begin{equation}{\mathcal
L}_{\rm mc} = D\vartheta^\mu\wedge\tau_\mu +
R^{\alpha\beta}\wedge\rho_{\alpha\beta} - \Lambda_{\rm
mc}(\vartheta,\tau,\rho);\label{generalL}
\end{equation}
 one of the associated
``covariant-symplectic'' boundary terms is\cite{CN9}
\begin{equation}
{\mathcal B}_{\rm mc}(N) = {\mathcal B}_\vartheta(N) +
\Delta\Gamma^{\alpha\beta}\wedge i_N\rho_{\alpha\beta} +
{\buildrel\scriptstyle\circ\over{D}}_\beta{\buildrel\scriptstyle\circ\over
{N}}{}^\alpha \Delta\rho_\alpha{}^\beta. \label{Bgen}
\end{equation}
The first order field equations are obtained by independent
variation with respect to $\vartheta$, $\tau$, $\Gamma$, and
$\rho$.  Extra Lagrange multiplier fields, as we shall see, can be
included to get the various special types of geometries.

\subsection{Riemannian General Relativity}

Einstein's general relativity, in its orthodox Riemannian
representation, can be obtained from the choice
\begin{equation}
\Lambda_{\rm mc} = \Lambda_{\rm
GR}:=V^{\alpha\beta}\wedge(\rho_{\alpha\beta}
-\eta_{\alpha\beta}).\label{LambdaGR}
\end{equation}
(Here we used Trautman's  dual basis:
$\eta^{\alpha\beta\cdots}:=*(\vartheta^\alpha\wedge\vartheta^\beta\wedge
\cdots)$, which is sometimes convenient.) The quantity
$V^{\alpha\beta}$ is a Lagrange multiplier field; its variation
yields $\rho_{\alpha\beta}=\eta_{\alpha\beta}$. Variation of
(\ref{generalL}), with the specification (\ref{LambdaGR}), with
respect to $\tau_\mu$ (since $\Lambda_{\rm GR}$ is independent of
$\tau_\mu$) simply yields the Riemannian connection's vanishing
torsion constraint: $D\vartheta^\mu=0$. One consequence is that
$D\eta_{\alpha\beta}=D\vartheta^\gamma\wedge\eta_{\alpha\beta\gamma}=0$;
from which it follows that the (vacuum)
$\delta\Gamma^{\alpha\beta}$ equation,
\begin{equation}
\vartheta_{[\beta}\wedge\tau_{\alpha]}+D\eta_{\alpha\beta}=0,
\end{equation}
yields $\tau_\mu=0$.  The general Hamiltonian boundary term
(\ref{Bgen}) reduces then to
\begin{equation}
{\mathcal B}_{GR}(N)=\Delta\Gamma^{\alpha\beta}\wedge
i_N\eta_{\alpha\beta} +
{\buildrel\scriptstyle\circ\over{D}}_\beta{\buildrel\scriptstyle\circ\over
{N}}{}^\alpha \Delta\eta_\alpha{}^\beta. \label{Bgr}
\end{equation}
This quasilocal expression was independently found from a
different perspective.\cite{KBL}  Asymptotically it agrees with
accepted expressions for energy-momentum, angular momentum and the
center-of-mass moment.\cite{RT,BO,Sza,MTW,ADM}

\subsection{Teleparallel Gravity and GR${}_{||}$ }

For teleparallel theories, in addition to a tetrad, we introduce a
new (metric compatible but not symmetric) connection, $\bar D$,
and force it to be teleparallel. This is accomplished by simply
choosing the first order Lagrangian potential to be independent of
$\rho$. (The potential $\Lambda$ is then formally like that for a
tetrad theory.)
\begin{equation}
{\mathcal L}_{||} = \bar D\vartheta^\mu\wedge\tau_\mu+\bar
R^{\alpha\beta}\wedge\rho_{\alpha\beta}-\Lambda_{||}(\vartheta,\tau).\label{teleL}
\end{equation}
Now variation with respect to $\rho$ yields the teleparallel
condition $\bar R^{\alpha\beta}=0$.

Since the curvature vanishes, parallel transport is path
independent. Hence one can choose a frame at one point and then
uniquely transport it to every other point. This preferred {\it
orthoteleparallel} (OT) frame is unique up to global (i.e. rigid,
constant) Lorentz transformations. (Note: teleparallel physics has
a preferred frame, it  does not have local Lorentz frame
invariance.) In the OT frame the connection coefficients vanish,
consequently the equations reduce to the tetrad form.  However,
the quasilocal expressions {\it do not} reduce to the tetrad form.

Comparing the Hamiltonian boundary term (\ref{Bgen}) with that of
the tetrad case (\ref{Btet}), we note extra terms involving
$\rho$, the connection's canonically conjugate momentum field. We
now show that for teleparallel theories this field generally
cannot be made to vanish, unlike the connection in an OT frame.
The quantity $\rho$ must satisfy the equation obtained by
variation with respect to $\bar\Gamma$:
\begin{equation}
\vartheta_{[\beta}\wedge\tau_{\alpha]}+\bar
D\rho_{\alpha\beta}=0.\label{rhoeq}
\end{equation}
From an orthodox $\Lambda_{||}$, quadratic in $\tau$, the
$\delta\tau$ equation gives a $\tau$ linear in $\bar D\vartheta$;
hence for teleparallel theories, $\tau$ is generally
non-vanishing. Consequently (\ref{rhoeq}) shows that $\rho$ is
generally non-vanishing.   Thus, although there is always a frame
in which the teleparallel connection coefficients vanish, the
connection still indirectly makes an important contribution
through its non-vanishing conjugate momentum field.  The relation
(\ref{rhoeq}) also shows another surprising feature: for all
teleparallel theories $\rho$ has the gauge freedom
\begin{equation}
\rho_{\alpha\beta} \to \rho_{\alpha\beta}+\bar
D\chi_{\alpha\beta},
\end{equation}
since $\bar D^2\chi\simeq \bar R\wedge \chi=0$.

In order to obtain GR${}_{||}$, the teleparallel equivalent  of
GR, we may take
\begin{equation}
\Lambda_{{\rm GR}_{||}} =
V^\mu\wedge(\tau_\mu-\kappa^{\alpha\beta}\wedge\eta_{\alpha\beta\mu})
-\kappa^\alpha{}_\gamma\wedge\kappa^{\gamma\beta}\wedge\eta_{\alpha\beta},
\label{TEGR}
\end{equation}
where the auxiliary one-form field $\kappa$ is used like a
Lagrange multiplier. Variation with respect to $\kappa$ yields
\begin{equation}
V^\mu\wedge\eta_{\alpha\beta\mu}
+\kappa^\lambda{}_\alpha\wedge\eta_{\lambda\beta}
+\kappa^\lambda{}_\beta\wedge\eta_{\alpha\lambda}=0.
\end{equation}
Incorporating the $\tau$ variation result, $V^\mu=\bar
D\vartheta^\mu$, the first term becomes $\bar
D\eta_{\alpha\beta}$. Consequently we infer that the tensorial
one-form $\kappa$ is just $
\kappa^\alpha{}_\beta=\Gamma^\alpha{}_\beta-\bar\Gamma^\alpha{}_\beta
$: the difference between the Levi-Civita connection coefficients
(which can be thought of as a certain function of the teleparallel
variables) and the teleparallel connection coefficients. Variation
with respect to $V^\mu$ gives
\begin{equation}
\tau_\mu=\kappa^{\alpha\beta}\wedge\eta_{\alpha\beta\mu}.
\end{equation}
We then find a solution to (\ref{rhoeq}):
$\rho_{\alpha\beta}=\eta_{\alpha\beta}$ (we fix the
$\chi_{\alpha\beta}$ gauge freedom with this choice).

Now we find the GR${}_{||}$ quasilocal boundary expression. The
teleparallel connection $\bar\Gamma$ is flat. We choose the
reference to be flat Minkowski space; specifically we take
\begin{equation}{{\buildrel\scriptstyle\circ\over{\bar\Gamma}}{}^{\alpha\beta}
=\buildrel\scriptstyle\circ\over\Gamma}{}^{\alpha\beta},\label{refcons}
\end{equation}
from which it follows that
${\buildrel\scriptstyle\circ\over\kappa}=0$, consequently
${\buildrel\scriptstyle\circ\over\tau}=0$. From these results we
find
\begin{equation}
{\mathcal B}_{{\rm GR}_{||}}(N)=i_N\vartheta^\mu
\kappa^{\alpha\beta}\wedge\eta_{\alpha\beta\mu}+\Delta{\bar\Gamma}{}^{\alpha\beta}\wedge
i_N\eta_{\alpha\beta}+{\buildrel\scriptstyle\circ\over{D}}{}^\beta{\buildrel\scriptstyle\circ\over
{N}}{}^\alpha \Delta\eta_{\alpha\beta}. \label{Btel}
\end{equation}
Our ${\mathcal B}_{{\rm GR}_{||}}$ expression asymptotically
agrees with the expression found by Blagojevi{\'c} and
Vasili{\'c}\cite{BV} for the teleparallel total quantities at
spatial infinity.

\section{The Quasilocal
Expressions}\label{sec:tettelepqle}

In this section we summarize the various quasilocal expressions in
a convenient common notation: the Riemannian variables.

We first observe that, because $K=\Gamma-\bar\Gamma$ and the fact
(\ref{refcons}) that we choose $\Gamma$ and $\bar\Gamma$ to have
the same reference values, the two expressions, ${\mathcal
B}_{{\rm GR}_{||}}$ (\ref{Btel}) and ${\mathcal B}_{{\rm GR}}$
(\ref{Bgr}), turn out to be {\it equivalent}. So they have the
{\it exactly} the same value for all quasilocal quantities.

With this in mind the various quasilocal expressions reduce to
\begin{eqnarray}
{\mathcal B}_{\rm tet}(N)&=&\Gamma^{\alpha\beta}\wedge i_N
\eta_{\alpha\beta}, \\
%\end{equation}
%\begin{equation}
{\mathcal B}_\vartheta(N)&=&\Delta\Gamma^{\alpha\beta}\wedge
i_N\eta_{\alpha\beta},\\
%\end{equation}
%\begin{eqnarray}
{\mathcal B}_{{\rm GR}_{||}}(N)\equiv {\mathcal B}_{\rm
GR}(N)&=&\Delta\Gamma^{\alpha\beta}\wedge i_N\eta_{\alpha\beta}+
{\buildrel\scriptstyle\circ\over{D}}{}^\beta{\buildrel\scriptstyle\circ\over
{N}}{}^\alpha \Delta\eta_{\alpha\beta}.\label{BgrBgrp}
\end{eqnarray}

Note that each of these quasilocal expressions has a ``Freud
type'' term, linear in $N$. Essentially the same Freud type term
as in ${\mathcal B}_{tet}$ has been derived by many investigators
including M{\o}ller,\cite{Mol} Nester,\cite{Ne89}
Maluf\cite{Mal,Mal2} and  Pereira.\cite{AP,AGP}   Only the last
quasilocal relation (\ref{BgrBgrp}) has an additional ``Komar
like'' $DN$ term. The {\it main point} we wish to convey is that
this $DN$ term is quite important for angular momentum and
absolutely {\it essential} for the center-of-mass. We remark that
such a term has long been recognized in GR Hamiltonian
investigations as essential for the correct boundary conditions
and correct total quantities at spatial infinity.\cite{RT,BO,Sza}

\begin{table}[b]
\tbl{Quasilocal COM from different expressions}
{%\footnotesize
\begin{tabular}{ccccc}
\hline \rule [-4.5mm]{0mm}{10mm}&  & $\mathcal{B}_{\rm tet}$ &
$\mathcal{B}_\vartheta$ & $\mathcal{B}_{{\rm GR}_{||}} \simeq
  \mathcal{B}_{\rm GR}$\\
  \hline
  \rule [-4.5mm]{0mm}{10mm} Freud term & & $\frac{2}{3}M\vec{d}$ & $\frac{2}{3}M\vec{d}$ & $\frac{2}{3}M\vec{d}$ \\
%  \hline
  \rule [-4.5mm]{0mm}{10mm} Komar term & & --- & --- & $\frac{1}{3}M\vec{d}$ \\
%  \hline
  \rule [-4.5mm]{0mm}{10mm} Total & & $\frac{2}{3}M\vec{d}$ & $\frac{2}{3}M\vec{d}$ & $M\vec{d}$ \\
  \hline
\end{tabular} }
\end{table}

\section{Evaluation}\label{sec:evaluation}

Here we discuss testing the various quasilocal expressions,
especially on the asymptotic eccentric Schwarzschild geometry.
Asymptotically the displacement vector field should approach a
Killing vector of the asymptotic Minkowski space.  Consequently it
should have the Poincar{\'e} Lie algebra form
\begin{equation}
N^\mu=N^\mu_0+\lambda^\mu{}_\nu x^\nu, \qquad \hbox{where}\quad
\lambda^{\mu\nu}=\lambda^{[\mu\nu]}.
\end{equation}

The constant spacetime translation $N^\mu_0$ (leading to
energy-momentum) and the spatial rotations (leading to angular
momentum) were investigated for the tetrad-teleparallel theory by
Vu.\cite{Vu} For GR, the center-of-mass moment was investigated by
Meng.\cite{Mng,NMC} We have included their results in our tables.
Here, for the tetrad-teleparallel theory, we examine the remaining
quasilocal quantity, the tetrad-teleparallel COM.

Consider the asymptotic spherically symmetric frame
\begin{equation}
\vartheta^t=(1+\Phi)dt, \qquad \vartheta^k=(1-\Phi)dx^k, \quad
\hbox{with} \quad \Phi=-{M\over r}.
\end{equation}
To test the expressions for the center-of-mass moment, displace
the center:
\begin{equation}
{1\over r}\to {1\over |{\bf r}-{\bf d}|}={1\over
r}\sum_{l=0}^{l=\infty}\bigl({d\over r}\bigr)^l
P_l(\cos\theta)\simeq {1\over r}+ {dz\over r^3} + \dots \quad.
\end{equation}
Now evaluate the quasilocal expressions for $N^\mu$ an asymptotic
boost. This choice of displacement detects the COM.
Straightforward calculations, taking the asymptotic limit of the
integral over a constant $r$  2-sphere (using the obvious
Minkowski reference where necessary), yield the results for the
various quasilocal expressions which are presented in Table 1.
Note that ${\mathcal B}_{\rm tet}$ and ${\mathcal B}_{\vartheta}$
do not give the correct COM value, whereas ${\mathcal B}_{{\rm
GR}_{||}}$ and ${\mathcal B}_{\rm GR}$ do. (Actually it {\it is }
possible to obtain the desired value from ${\mathcal
B}_\vartheta$---but {\it not} from ${\mathcal B}_{{\rm
tet}}$---{\it if} one selects a certain non-trivial reference;
this can be inferred from a recent COM work in GR.\cite{NMC,BLP})
Table 2 summarizes the limitations of the various
tetrad-teleparallel quasilocal expressions.

\begin{table}[h]
\tbl{Limitations of the quasilocal expressions}
{%\footnotesize
\begin{tabular}{cccc}
 \hline
 \rule [-4.5mm]{0mm}{10mm} & EM & AM & COM \\
 \hline
 \rule [-4.5mm]{0mm}{10mm}$\mathcal{B}_{\rm tet}$ & Special frame & Special frame & Fail \\
 \rule [-4.5mm]{0mm}{10mm}$\mathcal{B}_{\vartheta}$ & General frame & Special frame & Special reference \\
 \rule [-4.5mm]{0mm}{10mm}$\mathcal{B}_{{\rm GR}_{||}}\simeq \mathcal{B}_{\rm
 GR}$ & General frame & General frame & General frame \\
  \hline
\end{tabular} }
\end{table}

\section{Summary and Conclusion}\label{sec:conclusion}

We considered an important neglected quantity: the {\it quasilocal
center-of-mass} in tetrad-teleparallel gravity. (We have
distinguished these two formalisms.  In our terminology the former
has only a tetrad field, the latter has a tetrad and a connection
which has vanishing curvature.) We used the covariant Hamiltonian
formalism, in which quasilocal quantities are given by the
Hamiltonian boundary term, along with the {\it covariant
symplectic} Hamiltonian boundary expressions.  As expected,
consideration of the COM not only gives the most restrictive
asymptotic  conditions on the variables but also gives strong
constraints on the acceptable expressions.

We found that the $DN$ terms, which are absent in ${\mathcal
B}_{\rm tet}$, the Hamiltonian boundary term of the tetrad theory,
play an important role in angular momentum and an essential role
in obtaining the correct center-of-mass moment. Consequently the
{\it tetrad} formulation {\it does not} give the correct COM.  The
covariant-symplectic tetrad expression, ${\mathcal
B}_{\vartheta}$, can give good values with special choices of
frame and reference. On the other hand the {\it teleparallel}
formulation {\it does give} the correct AM and  COM, quite
generally,  independent of the asymptotic choice of frame. We also
remark that MTW,\cite{MTW} Eq.~(20.9), gives the necessary
asymptotic form for all 10 Poincar{\'e} quasilocal quantities. We
found that only the expressions ${\mathcal B}_{\rm GR}$ and
${\mathcal B}_{{\rm GR}_{||}}$, via the $DN$ terms, have this
form.

Remarkably, one of our covariant-symplectic teleparallel
Hamiltonian boundary expressions turns out to be equivalent to one
of the covariant symplectic GR boundary expressions (our other
boundary expressions for GR and GR${}_{||}$ will differ from them
quasilocally but not asymptotically). This leads us to the
conclusion (which we will revise if a good frame gauge condition,
meeting all the desired criteria, is identified) that, contrary to
a common hope, for localization purposes, teleparallel GR is {\it
no} better (and no worse) than Einstein's GR.

\section*{Acknowledgements}
This work was supported by the National Science Council of the ROC
under the grants NSC92-2112-M-008-050 and NSC92-2119-M-008-024.

\end{document}